# The temperature dependent thermal vector potential in spinor Boltzmann equation

Zheng-Chuan Wang

The University of Chinese Academy of Sciences, P. O. Box 4588, Beijing 100049, China, wangzc@ucas.ac.cn

## Abstract

The thermal scalar and vector potential were introduced to investigate the thermal transport under a temperature gradient in terms of linear response theory[1,2]. However, the microscopic origin of these phenomenological thermal potentials hadn't been addressed clearly. In this manuscript, we try to derive a temperature dependent damping force based on the spinor Boltzmann equation (SBE), and relate it with the thermal $U(1)$ gauge potential, which is exactly the temperature dependent thermal scalar and vector potential. It is shown that the thermal potential originates from the scattering of conduction electrons and impurity or other scattering mechanisms. We also derive a temperature dependent inverse relaxation time, which depends on momentum, it is different from the usual constant relaxation time. We evaluate the temperature dependent damping force by an approximated analytical solution of SBE. The other physical observable such as charge current and spin current are also explored.

PACS: 72.25-b; 05.60.Gg; 05.70.ln; 72.10.Di

# I. Introduction

Classical Boltzmann equation is a powerful tool to explore the transport problems in dilute gas[3], normal metal, semiconductor or in the strongly interacting systems with quasi-particles[4]. However, it can not be used to study the transport procedure in quantum systems, because momentum $\vec{p}$ and the position $\vec{r}$ in the distribution function may not be determined simultaneously according to the uncertainty principle, so Wigner[5] proposed a quantum Boltzmann equation satisfied by the so-called Wigner distribution function which is defined as the quantum Wigner transformation on lesser Green function, and the quantum Boltzmann equation can also be derived from the formalism of non-equilibrium Green function theory[6,7]. If we include the spin freedom into the quantum Boltzmann equation, the spinor Boltzmann equation (SBE) can be obtained, which was proposed by Levy et al[8]. In 2004, Sheng et al. Also derived it from the non-equilibrium Green function theory at steady state[9], then we further extended it to the two-momentum case[10]. Till now, SBE had been widely utilized to investigate the spin-dependent transport in spintronics[11,12].

In 2023, we derived a temperature dependent damping force based on the classical Boltzmann equation, which originates from the scattering terms of transport particles and reservoir particles, a non-equilibrium temperature is then defined[13]. The temperature dependent damping force was also extended to quantum Boltzmann equation by Wang, where it was related to a thermal potential[14]. The thermal potential was firstly proposed by Luttinger who introduced a scalar function Ψ in 1964 to describe the thermal transport under a temperature gradient[1], then the thermal transport coefficient can be obtained by Kubo's linear response theory similar to the electric field case. Now, the thermal potential had been widely used to study the transport procedure of electron, magnon driven by the temperature gradient, even to the thermally induced torque[15-21]. In order to overcome the non-physical divergence in the thermal transport coefficient when $T \to 0$ in Luttinger's work[15,16,19], Tatara introduced a thermal vector potential in the Hamiltonian to replace the above scalar potential[2]. Whatever the thermal scalar potential or the vector potential were introduced phenomenologically, their microscopic origin are still unclear. In Ref.[14], we related the temperature dependent damping force with these thermal scalar and vector potentials, and expressed them by the distribution function in quantum Boltzmann equation. We addressed the microscopic origin of the thermal potential from the scattering mechanisms in this system. It is shown that the thermal scalar and vector potential are U(1) gauge potential[14].

In this manuscript, we will derive the temperature dependent damping force in SBE, and relate it with the thermal scalar and vector potentials, we will show that this thermal potential is a SU(2)×U(1) gauge potential, which is caused by the spin freedom of electron in SBE. This non- Abelian thermal potential originates from the scattering of electron and impurity in our system. This paper is organized as follows: in Sec.II, we try to derive the temperature dependent damping force and the thermal potential, in Sec.III, we solve the SBE by an approximated method and calculate the physical observable numerically, a summary and discussion is given in Sec.IV.

## II. Theoretical formalism

Consider the spin dependent electronic transport under an external electric field $\vec{E}$ through ferromagnet in spintronics device, which is governed by the SBE[8]:

$$\frac{\partial \hat{f}}{\partial t} + \hat{v} \cdot \frac{\partial \hat{f}}{\partial \vec{r}} - \frac{e\vec{E}}{\hbar} \cdot \frac{\partial \hat{f}}{\partial \vec{p}} + \frac{i}{\hbar}[\hat{\varepsilon}, \hat{f}] = -\left(\frac{\partial \hat{f}}{\partial t}\right)_{collision},$$
(1)

where $\hat{f}$ is the spinor distribution function which is a $2 \times 2$ matrix, $\hat{\varepsilon}(p) = \varepsilon(p)\hat{I} + \frac{1}{2}\vec{J}(p) \cdot \vec{\sigma}$ represents the spinor energy, in which $\vec{J}(p) = J(p)\vec{M}(x)$, $\vec{M}(x)$ is the unit vector of magnetization of ferromagnet. $\hat{v} = \frac{1}{\hbar}\frac{\partial \hat{\varepsilon}}{\partial \vec{p}} = \frac{1}{\hbar}\left(\frac{\partial \varepsilon}{\partial \vec{p}} + \frac{1}{2}\frac{\partial J}{\partial \vec{p}}\vec{M} \cdot \vec{\sigma}\right) = v\hat{I} + u_x\vec{M}(x) \cdot \vec{\sigma}$ denotes the spinor velocity. The scattering term $\left(\frac{\partial \hat{f}}{\partial t}\right)_{collision}$ in Eq.(1) can be expressed as[7]:

$\left(\frac{\partial \hat{f}}{\partial t}\right)_{collision} = 2\pi\{[\Sigma_t(p,x,t)G^<(p,x,t) - \Sigma^<(p,x,t)G_{\bar{t}}(p,x,t) - G_t(p,x,t)\}\Sigma^<(p,x,t) + G^<(p,x,t)\Sigma_{\bar{t}}(p,x,t)] - \frac{i\hbar}{2}[\Sigma_t, G^<] + \frac{i\hbar}{2}[\Sigma_t G_{\bar{t}}] - \frac{i\hbar}{2}[G_t, \Sigma^<] + \frac{i\hbar}{2}[G^<, \Sigma_{\bar{t}}]$,
(2)

where $\Sigma^>$ and $\Sigma^<$ are the greater and lesser self-energies, $G^>$ and $G^<$ are the greater and lesser Green functions, respectively, $G_{\bar{t}}$ and $G_t$ are the anti-time order and time order Green function, $G_t = G^r$-$G^<$ and $G_{\bar{t}} = G^> - G^r$. $\Sigma_{\bar{t}}$ and $\Sigma_t$ are the anti-time order and time order self-energies, $\Sigma_t = \Sigma^r$-$\Sigma^<$ and $\Sigma_{\bar{t}} = \Sigma^> - \Sigma^r$, in which $\Sigma^r$ is the retarded self-energy. The bracket$[A,B]$, i.e. $[\Sigma^<, ReG^r]$, is defined as $\frac{\partial A}{\partial \vec{r}}\frac{\partial B}{\partial \vec{p}} - \frac{\partial A}{\partial \vec{p}}\frac{\partial B}{\partial \vec{r}}$. If the conduction electron is scattered by impurity in the system, under the self-consistent Born approximation, the self-energy can be written as[7]:

$$\Sigma^{>,<}(p) = n_i \int \frac{d^3p'}{(2\pi)^3}(T_{pp'})^2 G^{>,<}(p),$$
(3)

where $T_{pp'} = V_{pp'} + \int \frac{d^3p'}{(2\pi)^3}\frac{V_{pp_1'}T_{p_1p'}}{\omega-\varepsilon_p+i\delta}$, $V_{pp'}$ is the scattering potential.

We first consider the zero order terms with respect to $\hbar$ in the right hand side of Eq.(2). Substituting the expression of self-energy Eq.(3) into them, we have

$2\pi[(n_i \int \frac{d^3p'}{(2\pi)^3}(T_{pp'})^2 G^<(p',x,t))G^<(p,x,t) -$

$(n_i \int \frac{d^3p'}{(2\pi)^3}(T_{pp'})^2 G^<(p',x,t))G^>(p,x,t) +$

$(n_i \int \frac{d^3p'}{(2\pi)^3}(T_{pp'})^2 G^<(p',x,t))G^r(p,x,t) -$

$G^r(p,x,t)(n_i \int \frac{d^3p'}{(2\pi)^3}(T_{pp'})^2 G^<(p',x,t)) -$

$G^<(p,x,t)(n_i \int \frac{d^3p'}{(2\pi)^3}(T_{pp'})^2 G^<(p',x,t)) +$

$G^<(p,x,t)(n_i \int \frac{d^3p'}{(2\pi)^3}(T_{pp'})^2 G^>(p',x,t))]$, (4)

where we have used the retarded self-energy $\Sigma^r(p) = n_i T_{pp}$. If the momentum transfer caused by the electron- impurity scattering is small compared with the momentum $\vec{p}$ of electron, we can make a Taylor series expansion on the Green function $G^<(p', x, t)$ and $G^>(p', x, t)$ around $\vec{p}$ in the above integrals, then Eq.(4) can be written as:

$2\pi[B\frac{\partial G^<}{\partial p}G^<(p,x,t) -$

$AG^<(p,x,t)G^>(p,x,t) - B\frac{\partial G^<}{\partial p}G^>(p,x,t) +$

$AG^<(p,x,t)G^r(p,x,t) + B\frac{\partial G^<}{\partial p}G^r(p,x,t) -$

$AG^r(p',x,t)G^<(p,x,t) - BG^r(p,x,t)\frac{\partial G^<}{\partial p} -$

$BG^<(p,x,t)\frac{\partial G^<}{\partial p} + AG^<(p,x,t)G^>(p,x,t) +$

$BG^<(p,x,t)\frac{\partial G^>}{\partial p}$,  (5)

where $A = n_i \int \frac{d^3p'}{(2\pi)^3}(T_{pp'})^2$, $B = n_i \int \frac{d^3p'}{(2\pi)^3}(T_{pp'})^2(p'-p)$. Since $G^< = i\hat{f}(p,x,t)$ and $G^> = i\hat{f}'(p,x,t)$, where $\hat{f}'(p,x,t)$ is the distribution function of hole, then the right hand side of Eq.(1) can be written as:

$2\pi[-B\frac{\partial \hat{f}}{\partial p}\hat{f}(p,x,t) + A\hat{f}(p,x,t)\hat{f}'(p,x,t) +$

$B\frac{\partial \hat{f}}{\partial p}\hat{f}'(p,x,t) + iA\hat{f}(p,x,t)G^r(p,x,t) +$

$iB\frac{\partial \hat{f}}{\partial p}G^r(p,x,t) - iAG^r(p,x,t)\hat{f}(p,x,t) -$

$iBG^r(p,x,t)\frac{\partial \hat{f}}{\partial p} + B\hat{f}(p,x,t)\frac{\partial \hat{f}}{\partial p} -$

$A\hat{f}(p,x,t)\hat{f}'(p,x,t) - B\hat{f}(p,x,t)\frac{\partial \hat{f}'}{\partial p}$.  (6)

If we neglect the first order terms of $\hbar$ in the right hand side of Eq.(2) and only keep the above zero order terms, then the SBE can be simplified as

$\frac{\partial \hat{f}}{\partial t} + \hat{v} \cdot \frac{\partial \hat{f}}{\partial \vec{x}} + (\frac{e\vec{E}}{2\hbar} + iBG^r(p,x,t) - B\hat{f} + \frac{J(p)}{4}\frac{\partial \vec{M}(x)}{\partial \vec{x}} \cdot$

$\vec{\sigma}) \cdot \frac{\partial \hat{f}}{\partial \vec{p}} + \frac{\partial \hat{f}}{\partial \vec{p}} \cdot (\frac{e\vec{E}}{2\hbar} - iBG^r(p,x,t) + B\hat{f} + \frac{J(p)}{4}\frac{\partial \vec{M}(x)}{\partial \vec{x}} \cdot$

$\vec{\sigma}) - \frac{J(p)}{2i\hbar}\vec{M}(x) \cdot \vec{\sigma}\hat{f} + \frac{J(p)}{2i\hbar}\hat{f}\vec{M}(x) \cdot \vec{\sigma} + iAG^r\hat{f} -$

$iA\hat{f}G^r = -B\hat{f}\frac{\partial \hat{f}'}{\partial \vec{p}} + B\frac{\partial \hat{f}}{\partial \vec{p}}\hat{f}'$.

(7)

We can see that the damping force $iBG^r(p,x,t) - B\hat{f}$ has appeared in Eq.(7). Since the spinor wavefunction is a 2×2 matrix, this force is a 2×2 matrix, which is spinor force. In fact, we can expand the spinor distribution by the complete basis of three Pauli matices and unit matrix based on the local equilibrium assumption as [14]:

$\hat{f}(p, x, t) = f^0(p, x) + \left(-\frac{\partial f^0}{\partial \varepsilon}\right)[f(p, x, t) + \vec{g}(p, x, t) \cdot \vec{\sigma}]$,  (8)

where $f^0(p, x) = \frac{1}{\exp\left[\frac{\varepsilon - \mu}{kT(x)}\right] + 1}$ is the local equilibrium distribution, $\mu$ is the chemical potential. The temperature $T(x)$ therein is position-dependent and its gradient will give rise to the thermal current and thermal spin current in the magnetic system. $f(p, x, t)$ and $\vec{g}(p, x, t)$ are the scalar distribution function and vector distribution function, respectively, which describe the non-equilibrium deviation of spinor distribution function from the local equilibrium distribution. Then the damping force can be expanded as:

$\hat{F} = (iBG_I^r - Bf)\hat{I} + (iB\vec{G}_\sigma^r - B\vec{g}) \cdot \vec{\sigma}$,  (9)

which is expanded by the matrices $\hat{I}$ and $\vec{\sigma}$, where $\hat{G}^r = G_I^r\hat{I} + \vec{G}_\sigma^r \cdot \vec{\sigma}$. Since we expand the spinor distribution function around the local equilibrium distribution function Eq.(8), the temperature $T(X)$ will appear in the local equilibrium distribution function, then appear in the damping force.Eq.(9), so the damping force is a temperature dependent thermal force, which can be related to the thermal scalar and vector potentials.

The scalar part $iBG_I^r - Bf$ of the damping force can be regarded as contributed by the

thermal scalar potential $\varphi$ with
$$-\nabla\varphi = iBG_I^r - Bf, \quad (10)$$
Then we can determine the thermal scalar potential from the above equation as:
$$\varphi = \int(-iBG_I^r + Bf)dx, \quad (11)$$
where the thermal scalar potential is expressed by the scalar distribution function and scalar retarded Green function, so it is temperature dependent, too.

The vector part $iB\vec{G}_\sigma^r - B\vec{g}$ of the damping force can be regarded as contributed by the thermal vector potential $\vec{A}$ with
$$e\vec{v} \times (\nabla \times \vec{A}_\sigma) = iB\vec{G}_\sigma^r - B\vec{g}, \quad (12)$$
We can determine the thermal vector potential $\vec{A}_\sigma$ from the above equation, which has a spinor indices and is expressed by the vector distribution and vector Green function, therefore is temperature dependent.

There exist gauge freedom for the above gauge potential. The damping force keep the gauge invariant under the U(1) gauge transformation. It should be noted that the spinor damping force originates from the scattering term which comes from the interaction of conduction electrons and the impurity, it is not the real weak interaction in particle physics.

Not only the zero order terms on the right hand side of Eq.(2) have contribution on the damping force, but also the first order terms in the collision term have contribution on the damping force, the latter even modify the velocity of electron and induce an anomalous velocity. Let us explore this issue. Similar to the Taylor series expansion in the zero order terms, the first order terms of $\hbar$ in Eq.(2), i.e. $-\frac{i\hbar}{2}[\Sigma_t, G^<]$, can be written as:

$$-\frac{i\hbar}{2}[\Sigma_t, G^<] = -\frac{i\hbar}{2}[-A\frac{\partial G^<}{\partial p}\frac{\partial G^<}{\partial x} - B\frac{\partial^2 G^<}{\partial^2 p}\frac{\partial G^<}{\partial x} +$$
$$A\frac{\partial G^<}{\partial x}\frac{\partial G^<}{\partial p} + B\frac{\partial^2 G^<}{\partial x\partial p}\frac{\partial G^<}{\partial p}], \quad (13)$$

where we have substituted the self-energy Eq.(3) into the above bracket. Similarly, we can perform the same operation on the other first order terms in Eq.(2), the total contribution for the first order terms of $\hbar$ in the right hand side of Eq.(2) are given in Appendix A. When we put them into the SBE and rearrange the form of $\frac{\partial \hat{f}}{\partial \vec{x}}$ term and $\frac{\partial \hat{f}}{\partial \vec{p}}$ term, we can obtain the SBE with the first order quantum corrections, which is shown in Appendix B.

We can see that the first order terms have contribution $-B\frac{\partial^2 \hat{f}}{\partial x \partial p} - iA\frac{\partial G^r}{\partial x}$ on the damping force, moreover, it also have contribution on the velocity of electron by an anomalous velocity as $i\frac{\partial(n_i T_{pp})}{\partial x} + A\frac{\partial \hat{f}}{\partial \vec{p}} + 2B\frac{\partial^2 \hat{f}}{\partial^2 p} + iA\frac{\partial G^r}{\partial p} - 3i\frac{\partial(n_i T_{pp})}{\partial p}$, which is similar to the anomalous velocity in anomalous Hall effect[19]. As shown in the above, the damping force corrections and anomalous velocity originate from the first order scattering terms of conduction electrons and impurities, which are temperature dependent, too.

The terms $-B\hat{f}\frac{\partial \hat{f}'}{\partial \vec{p}} + B\frac{\partial \hat{f}}{\partial \vec{p}}\hat{f}'$ on the right hand side of Eq.(7) correspond to the relaxation term $-\frac{\hat{f}-<\hat{f}>}{\tau}$. Compared them, we can obtain $\tau = -\frac{\hat{f}-<\hat{f}>}{(-B\hat{f}\frac{\partial \hat{f}'}{\partial \vec{p}} + B\frac{\partial \hat{f}}{\partial \vec{p}}\hat{f}')}$, which depends on the momentum, it is not a constant as some previous works, more important, it is also temperature dependent, we call it the damping relaxation time.

### III. The solution of temperature dependent SBE

In this section, we will try to solve the SBE (7) and evaluate the temperature dependent damping force and other physical observable. If we decompose the spinor distribution function as Eq.(8) and substitute it into SBE (7), we can obtain the equation for the scalar distribution function as:

$$\left[\frac{\partial}{\partial t} + \frac{p}{m}\frac{\partial}{\partial x} + eE\frac{\partial}{\partial p}\right]f(p,x,t) + \frac{J(p)}{2}\frac{\partial \vec{M}(x)}{\partial x} \cdot \frac{\partial \vec{g}}{\partial p} = B\frac{\partial f}{\partial p}f' + B\frac{\partial \vec{g}}{\partial p} \cdot \vec{g}' - B\frac{\partial f'}{\partial p}f - B\frac{\partial \vec{g}'}{\partial p} \cdot \vec{g}, \quad (14)$$

and the equation for the vector distribution function as:

$$\left[\frac{\partial}{\partial t} + \frac{p}{m}\frac{\partial}{\partial x} + eE\frac{\partial}{\partial p}\right]\vec{g}(p,x,t) - 2B\vec{G}_g^r \times \frac{\partial \vec{g}}{\partial p} - 2iB\vec{g} \times \frac{\partial \vec{g}}{\partial p} - \frac{J}{2}\frac{\partial \vec{M}(x)}{\partial x}\frac{\partial f}{\partial p} - \frac{J}{\hbar}\left(\vec{M}(x) \times \vec{g}(p,x,t)\right) - 2A\vec{G}_g^r \times \vec{g} = -B\frac{\partial f}{\partial p}\vec{g}' + B\frac{\partial \vec{g}}{\partial p}f' + iB\frac{\partial \vec{g}}{\partial p} \times \vec{g}' - Bf\frac{\partial \vec{g}'}{\partial p} - B\vec{g}\frac{\partial f'}{\partial p} - iB\vec{g} \times \frac{\partial \vec{g}'}{\partial p}. \quad (15)$$

It is shown that the above two equations are coupled together, we must solve them simultaneously. It is very difficult to obtain the analytical solutions for Eq.(14) and (15), because they are complicated integro-differential equation group, but we can try to solve them by some reasonable approximations. For simplicity, we only consider the one-dimensional spin dependent transport through a ferromagnet. Let us start from the equation satisfied by the x-component of vector distribution function, it is

$$\frac{\partial}{\partial t}g_x + \frac{p}{m}\frac{\partial}{\partial x}g_x + (eE + Bf')\frac{\partial g_x}{\partial p} + 2B\frac{\partial f}{\partial p}g_x = F_1, \quad (16)$$

where $F_1 = B\frac{\partial f}{\partial p}g'_x + 2B(G^r_{gy}\frac{\partial g_z}{\partial p} - G^r_{gz}\frac{\partial g_y}{\partial p}) + 2iB(g_y\frac{\partial g_z}{\partial p} - g_z\frac{\partial g_y}{\partial p}) - \frac{J(p)}{2}\frac{\partial M_x}{\partial x}\frac{\partial f}{\partial p} + \frac{J(p)}{\hbar}(M_y g_z - M_z g_y) + 2A(G^r_{gy}g_z - G^r_{gz}g_y) + iB(g'_z\frac{\partial g_y}{\partial p} - g'_y\frac{\partial g_z}{\partial p}) - Bf\frac{\partial g'_x}{\partial p} - iB(g_y\frac{\partial g'_z}{\partial p} - g_z\frac{\partial g'_y}{\partial p})$. If we assume that $g_x(p,x,t)$ has the form of separating variable:

$$g_x(p,x,t) = g(p)g(x)g(t), \quad (17)$$

Since the system will arrive at the steady state after a relaxation time τ, we usually study the transport at steady state in which $\frac{\partial}{\partial t}g_x = 0$, by use of Eq.(17) we have

$$D_1\frac{\partial g(p)}{\partial p} + D_2 g(p) = F_1, \quad (18)$$

where $D_1 = (e\frac{\partial U}{\partial x} - Bf')g(x)g(t)$ and $D_2 = g(x)\frac{\partial g(t)}{\partial t} + \frac{p}{m}g(t)\frac{\partial g(x)}{\partial x} + B\frac{\partial f'}{\partial p}g(x)g(t)$.

Eq(18) has the following formal solution

$$g(p) = e^{-\int \frac{D_2}{D_1}dp}\left(\int \frac{F_1}{D_1}e^{\int \frac{D_2}{D_1}dp}dp\right), \quad (19)$$

which is still expressed by some unknown functions $g(x)$ and $g(t)$, we must adopt some approximations as starting point. At steady state, the scalar distribution function can be approximately adopted as the equilibrium function $f^0(p,X)$, the decay of $g(t)$ can be simplified as $e^{-\frac{t}{\tau}}$. Meanwhile, the spin density of electron-- the spin accumulation $\vec{m} = \int \vec{g}dp$ will have the same direction with the magnetization of background ferromagnet. If we set the magnetization of ferromagnet as (1,0,0), then $g_y$ and $g_z$ will become to zero at steady state, and only $g_x$ is not zero, then g(p)can be obtained approximately. Then we try to solve $g(t)$, from Eq.(16), it satisfies

$$E_1\frac{\partial g(t)}{\partial t} + E_2 g(t) = F_1, \quad (20)$$

which has a formal solution

$$g(t) = e^{-\int \frac{E_2}{E_1}dt}(\int \frac{F_1}{E_1} e^{\int \frac{E_2}{E_1}dt} dt), \qquad (21)$$

where $E_1 = g(x)g(p)$ and $E_2 = (e\frac{\partial U}{\partial x} - Bf')g(x)\frac{\partial g(p)}{\partial p} + \frac{p}{m} g(p) \frac{\partial g(x)}{\partial x} + B\frac{\partial f'}{\partial p} g(x)g(p)$. Substituting the expression (19) of g(p) into the above expression, and $g(x)$ adopting the similar approximations as in Eq.(19), we can obtain the time dependent g(t). Similarly, the equation satisfied by $g(x)$ as:

$$G_1 \frac{\partial g(x)}{\partial x} + G_2 g(x) = F_1, \qquad (22)$$

with a solution

$$g(x) = e^{-\int \frac{G_2}{G_1}dx}(\int \frac{F_1}{G_1} e^{\int \frac{G_2}{G_1}dx} dx, \qquad (23)$$

where $G_1 = \frac{p}{m} g(p)g(t)$ and $G_2 = g(p)\frac{\partial g(t)}{\partial t} + (e\frac{\partial U}{\partial x} - Bf')g(t)\frac{\partial g(p)}{\partial p} + B\frac{\partial f'}{\partial p} g(t)g(p)$.

Substituting the above g(p) and g(t) into the Eq.(23), we can obtain the x-component of vector function $g_x(p,x,t) = g(p)g(x)g(t)$, so we finally arrive at the solution at the first order approximation, then we can make further iteration to get higher order solutions. Similarly, the y-component of vector distribution function satisfies:

$$\frac{\partial}{\partial t} g_y + \frac{p}{m} \frac{\partial}{\partial x} g_y + (eE + Bf')\frac{\partial g_y}{\partial p} + 2B \frac{\partial f}{\partial p} g_y = F_2, \qquad (24)$$

when we separate $g_y(p,x,t)$ as $G(p)G(x)G(t)$, analogous to the way for $g_x(p,x,t)$ in the above, at steady state we have

$$H_1 \frac{\partial G(p)}{\partial p} + H_2 G(p) = F_2, \qquad (25)$$

where $g_z$ is still adopted as zero, $g_x$ is adopted as Eq.(17), $H_1 = (e\frac{\partial U}{\partial x} - Bf')G(x)G(t)$,

$H_2 = G(x)\frac{\partial G(t)}{\partial t} + \frac{p}{m} G(t)\frac{\partial G(x)}{\partial x} + B\frac{\partial f'}{\partial p} G(x)G(t)$

and $F_2 = B\frac{\partial f}{\partial p} g'_y + 2B(G^r_{gz}\frac{\partial g_x}{\partial p} - G^r_{gx}\frac{\partial g_z}{\partial p}) + 2iB(g_z \frac{\partial g_x}{\partial p} - g_x \frac{\partial g_z}{\partial p}) - \frac{J(p)}{2}\frac{\partial M_y}{\partial x}\frac{\partial f}{\partial p} + \frac{J(p)}{\hbar}(M_z g_x - M_x g_z) + 2A(G^r_{gz}g_x - G^r_{gx}g_z) + iB(g'_z\frac{\partial g_x}{\partial p} - g'_x\frac{\partial g_z}{\partial p}) - Bf\frac{\partial g'_y}{\partial p} - iB(g_z\frac{\partial g'_x}{\partial p} - g_x\frac{\partial g'_z}{\partial p})$. In the above, we have approximated $G(t)$ as $e^{-\frac{t}{\tau}}$, then

$$G(p) = e^{-\int \frac{H_2}{H_1}dp}(\int \frac{F_2}{H_1} e^{\int \frac{H_2}{H_1}dp} dp, \qquad (26)$$

Similarly, we have

$$G(t) = e^{-\int \frac{K_2}{K_1}dt}(\int \frac{F_2}{K_1} e^{\int \frac{K_2}{K_1}dt} dt, \qquad (27)$$

with $K_1 = G(x)G(P)$, $K_2 = \frac{p}{m} G(p)\frac{\partial G(x)}{\partial x} + (e\frac{\partial U}{\partial x} - Bf')G(x)\frac{\partial G(p)}{\partial p} + B\frac{\partial f'}{\partial p} G(x)G(p)$, and

$$G(x) = e^{-\int \frac{L_2}{L_1}dx}(\int \frac{F_2}{L_1} e^{\int \frac{L_2}{L_1}dx} dx, \qquad (28)$$

with $L_1 = \frac{p}{m} G(p)G(t)\frac{\partial G(x)}{\partial x}$ and $L_2 = G(p)\frac{\partial G(t)}{\partial t} + (e\frac{\partial U}{\partial x} - Bf')G(t)\frac{\partial G(p)}{\partial p} + B\frac{\partial f'}{\partial p} G(t)G(p)$, finally we can obtain the approximated solution of $g_y(p,x,t) = G(p)G(x)G(t)$. Repeat the same procedure for $g_z(p,x,t) = g'(p)g'(x)g'(t)$, where $g_x$ and $g_y$ are substituted by the above solutions obtained, we can get the z-component of vector distribution function, in which

$$g'(p) = e^{-\int \frac{N_2}{N_1}dp}(\int \frac{F_3}{N_1} e^{\int \frac{N_2}{N_1}dp} dp, \qquad (29)$$

with $N_1 = (e\frac{\partial U}{\partial x} - Bf')g'(x)g'(t)$ and $N_2 = g'(x)\frac{\partial g(t)}{\partial t} + \frac{p}{m} g'(t)\frac{\partial g'(x)}{\partial x} + B\frac{\partial f'}{\partial p} g'(t)g'(x)$

and $F_3 = B\frac{\partial f}{\partial p}g'_z + 2B(G^r_{gx}\frac{\partial g_y}{\partial p} - G^r_{gy}\frac{\partial g_x}{\partial p}) + 2iB(g_x\frac{\partial g_y}{\partial p} - g_y\frac{\partial g_x}{\partial p}) - \frac{J(p)}{2}\frac{\partial M_z}{\partial x}\frac{\partial f}{\partial p} + \frac{J(p)}{\hbar}(M_x\ g_y - M_y\ g_x) + 2A(G^r_{gx}g_y - G^r_{gy}g_x) + iB(g'_x\frac{\partial g_y}{\partial p} - g'_y\frac{\partial g_x}{\partial p}) - Bf\frac{\partial g'_z}{\partial p} - iB(g_x\frac{\partial g'_y}{\partial p} - g_y\frac{\partial g'_x}{\partial p})$, and

$$g'(t) = e^{-\int \frac{M_2}{M_1}dt}(\int \frac{F_3}{M_1}e^{\int \frac{M_2}{M_1}dt}dt, \qquad (30)$$

with $M_1 = g'(p)g'(x)$ and $M_2 = \frac{p}{m}g'(p)\frac{\partial g'(x)}{\partial x} + (e\frac{\partial U}{\partial x} - Bf')g(x)\frac{\partial g'(p)}{\partial p} + B\frac{\partial f'}{\partial p}g'(p)g'(x)$, and

$$g'(x) = e^{-\int \frac{P_2}{P_1}dx}(\int \frac{F_3}{P_1}e^{\int \frac{P_2}{P_1}dx}dx, \qquad (31)$$

with $P_1 = \frac{p}{m}g'(p)g'(t)$ and $P_2 = g'(p)\frac{\partial g'(t)}{\partial t} + (e\frac{\partial U}{\partial x} - Bf')g'(t)\frac{\partial g'(p)}{\partial p} + B\frac{\partial f'}{\partial p}g'(t)g'(p)$.

The similar procedure can be performed for the scalar distribution function $f(p,x,t) = F(p)F(x)F(t)$, where

$$F(p) = e^{-\int \frac{Q_2}{Q_1}dp}(\int \frac{F_4}{Q_1}e^{\int \frac{Q_2}{Q_1}dp}dp, \qquad (32)$$

with $Q_1 = (e\frac{\partial U}{\partial x} - Bf')F(x)F(t)$, $Q_2 = F(x)\frac{\partial F(t)}{\partial t} + \frac{p}{m}F(t)\frac{\partial F(x)}{\partial x}$, and $F_4 = -\frac{J(p)}{2}(\frac{\partial M_x}{\partial x}\frac{\partial g_x}{\partial p} + \frac{\partial M_y}{\partial x}\frac{\partial g_y}{\partial p} + \frac{\partial M_z}{\partial x}\frac{\partial g_z}{\partial p}) - Bf\frac{\partial f'}{\partial p} + B(\frac{\partial g_x}{\partial p}g'_x + \frac{\partial g_y}{\partial p}g'_y + \frac{\partial g_z}{\partial p}g'_z) - B(\frac{\partial g'_x}{\partial p}g_x + \frac{\partial g'_y}{\partial p}g_y + \frac{\partial g'_z}{\partial p}g_z)$, and

$$F(t) = e^{-\int \frac{S_2}{S_1}dt}(\int \frac{F_5}{S_1}e^{\int \frac{S_2}{S_1}dt}dt, \qquad (33)$$

with $S_1 = F(x)F(p)$ and $N_2 = \frac{p}{m}F(p)\frac{\partial F(x)}{\partial x} + (e\frac{\partial U}{\partial x} - Bf')F(x)\frac{\partial F(p)}{\partial p}$, and

$$F(x) = e^{-\int \frac{R_2}{R_1}dx}(\int \frac{F_5}{R_1}e^{\int \frac{R_2}{R_1}dx}dx, \qquad (34)$$

with $R_1 = \frac{p}{m}F(t)F(p)$ and $R_2 = F(p)\frac{\partial F(t)}{\partial t} + (e\frac{\partial U}{\partial x} - Bf')F(t)\frac{\partial F(p)}{\partial p}$,

where $g_x$, $g_y$ and $g_z$ are substituted by the above solutions we get, we finally obtain the scalar distribution function.

After we obtain the approximated solution of SBE, we can calculate the charge density and charge current as $\rho(x,t) = e\int f(p,x,t)dp$ and $\vec{j}(x,t) = e\int \vec{v}f(p,x,t)dp$, the spin accumulation and spin current as $\vec{m}(x,t) = e\int \vec{g}(p,x,t)dp$ and $\vec{j}_m(x,t) = e\int \vec{v}\vec{g}(p,x,t)dp$, respectively.

As an example, we investigate the spin-polarized transport through a one-dimensional ferromagnet. The length of the system is chosen as 10nm, an external electric field $E = 0.01\mu V/nm$ is applied to the system, the magnetization in the ferromagnet is chosen as $\vec{M} = (1,0,0)$. In Fig.1 and Fig.2, we plot the scalar part and vector part of damping force as a function of position. Since the damping force Eq.(9) is a spinor, it has four components, one scalar component and three vector components. The scalar component of damping force is drawn in Fig.1, the figures for three components of vector damping force are similar, we only draw the z-component of vector damping force in Fig.2. Whatever in Fig.1 or Fig.2, the curves vary with position obviously, all of them depend on the temperature, where we have chosen the temperature as 300K. The

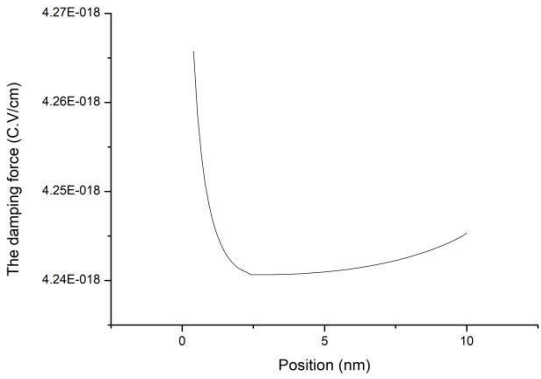

Fig1. The scalar damping force vs position

charge current density as a

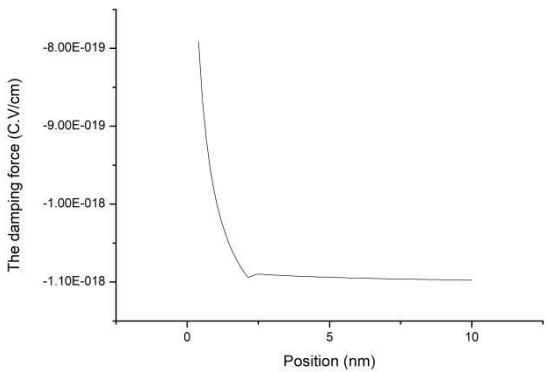

Fig.2 The z-component of damping force vs position

function of position is plotted in Fig.3, it is driven by the external electric field. The charge current density is position dependent, not a constant, the current density at the left and right boudaries are different, because the current is scattered by the electron- impurity interaction, the charge density and current density still satisfy the charge continuity equation. In Fig.4, we draw the x-component of spin current, it increases with position. The variation of spin current is caused by the spin transfer torque, the local magnetization of background ferromagnet will have influence on the spin current by torque which is demonstrate in Fig.5, the spin current and spin torque obey the spin diffusion equation. We only draw the z-component of spin transfer torque in Fig.5, the y-component is similar to the z-component, the x-component is zero, because the magnetization of background ferromagnet is chosen along x-axis, the torque is defined by the cross product of $\vec{M} \times \vec{m}(x)$, where $\vec{m}(x)$ is the spin accumulation. The thermal current density as a function of position is shown in Fig.6, it has a similar shape with the charge current, because the heat current is carried by the conduction electrons in our study, we haven't considered the heat current carried by the phonon in this manuscript.

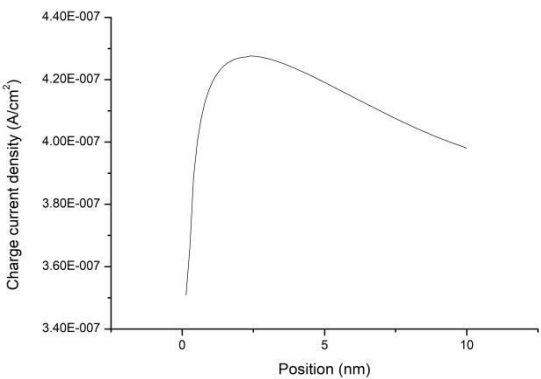

Fig.3 The charge current density vs position

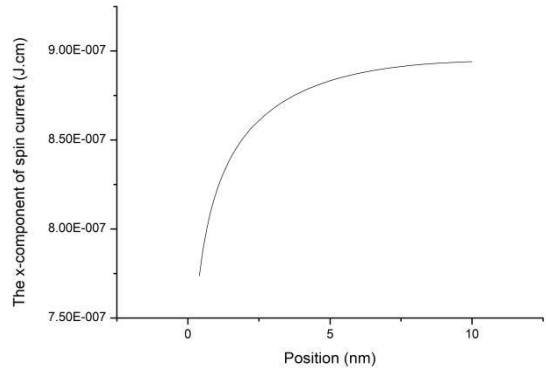

Fig.4 The x-component of spin current vs position

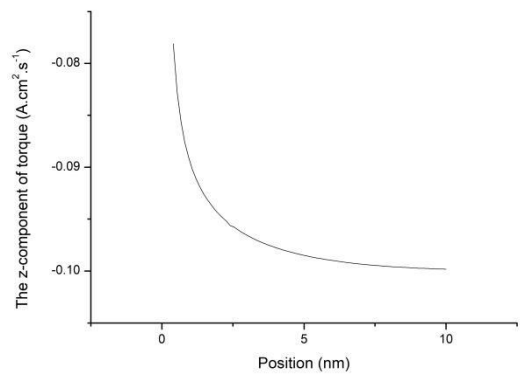

Fig.5 The z-component of spin transfer torque vs position

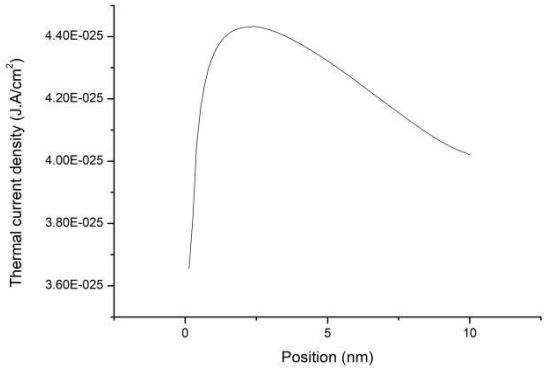

Fig.6 The thermal current density vs position

## IV. Summary and discussions

In this manuscript, we derive the temperature dependent damping force and the inverse relaxation time based on SBE. The spinor damping force is a 2×2 matrix, we relate it with the thermal U(1) gauge potential, and show it keep the gauge invariant. The first order terms of $\hbar$ in the scattering terms of SBE also contribute a anomalous velocity to the velocity of conduction electrons, which is similar to the anomalous Hall effect. Whatever the damping force, the inverse relaxation time or the anomalous velocity originate from the scattering of conduction electron with impurity. To evaluate the physical observable, we also propose an approximated analytical method to solve the complicated integro- differential equations satisfied by the scalar and vector distribution functions, which avoids the time consuming numerical calculation to solve the SBE. The physical observable such as damping force, charge current and spin current et al. are shown in Fig.1-Fig.6, respectively. It should be pointed out that we only consider the scattering of electron- impurity in our study, the other scattering mechanisms, i.e., the scattering of electron- phonon and electron- magnon should also be considered in the self-energy, they will contribute to the temperature dependent damping force and the thermal gauge potential, too, we leave it for future exploration.

## Acknowledgments

This study is supported by the National Key R&D Program of China (Grant No. 2022YFA1402703).

## Data Availability Statement

Data sets generated during the current the study are available from the corresponding author on reasonable request.

## Additional information

Competing interest statement: The authors declare that they have no competing interests.

## Appendix A:

The first order terms of $\hbar$ on the right hand side of Eq.(1) is:

$$-\frac{i\hbar}{2}[(\hat{v} + i\frac{\partial(n_i T_{pp})}{\partial x} + A\frac{\partial \hat{f}}{\partial \vec{p}} + 2B\frac{\partial^2 \hat{f}}{\partial^2 p} + iA\frac{\partial G^r}{\partial p}$$
$$- 3i\frac{\partial(n_i T_{pp})}{\partial p}) \cdot \frac{\partial \hat{f}}{\partial \vec{x}} + (-B\frac{\partial^2 \hat{f}}{\partial x \partial p}$$
$$- iA\frac{\partial G^r}{\partial x}) \cdot \frac{\partial \hat{f}}{\partial \vec{p}} + \frac{\partial \hat{f}}{\partial \vec{p}} \cdot (-iA\frac{\partial G^r}{\partial x}$$
$$- B\frac{\partial^2 \hat{f}}{\partial x \partial p}) + i\frac{\partial(n_i T_{pp})}{\partial x}\frac{\partial \hat{f}'}{\partial p}$$
$$+ i\frac{\partial(n_i T_{pp})}{\partial x}\frac{\partial \hat{f}}{\partial p} - \frac{\partial(n_i T_{pp})}{\partial x}\frac{\partial G^r}{\partial p}$$
$$+ B\frac{\partial^2 \hat{f}}{\partial x \partial p}\frac{\partial \hat{f}'}{\partial x} - iA\frac{\partial \hat{f}}{\partial x}\frac{\partial G^r}{\partial x}$$
$$+ iB\frac{\partial G^r}{\partial p}\frac{\partial^2 \hat{f}}{\partial x \partial p} - iB\frac{\partial^2 \hat{f}}{\partial^2 p}\frac{\partial G^r}{\partial x}$$
$$- iB\frac{\partial G^r}{\partial x}\frac{\partial^2 \hat{f}}{\partial^2 p} + B\frac{\partial \hat{f}}{\partial x}\frac{\partial^2 \hat{f}}{\partial^2 p}$$
$$- A\frac{\partial \hat{f}}{\partial x}\frac{\partial \hat{f}'}{\partial p} + iB\frac{\partial^2 \hat{f}'}{\partial x \partial p} + iB\frac{\partial^2 \widehat{f'}}{\partial x \partial p}$$
$$- \frac{\partial(n_i T_{pp})}{\partial p}\frac{\partial G^r}{\partial x} - iB\frac{\partial^2 \hat{f}}{\partial x \partial p}\frac{\partial G^r}{\partial x}]$$

The SBE with the first order quantum corrections is:

$$\frac{\partial \hat{f}}{\partial t} + (\hat{v} + i\frac{\partial(n_i T_{pp})}{\partial x} + A\frac{\partial \hat{f}}{\partial \vec{p}} + 2B\frac{\partial^2 \hat{f}}{\partial^2 p} + iA\frac{\partial G^r}{\partial p} -$$
$$3i\frac{\partial(n_i T_{pp})}{\partial p}) \cdot \frac{\partial \hat{f}}{\partial \vec{x}} + (\frac{e\vec{E}}{2\hbar} + iBG^r(p,x,t) - B\hat{f} +$$
$$\frac{J(p)}{4}\frac{\partial \vec{M}(x)}{\partial \vec{x}} \cdot \vec{\sigma} - B\frac{\partial^2 \hat{f}}{\partial x \partial p} - iA\frac{\partial G^r}{\partial x}) \cdot \frac{\partial \hat{f}}{\partial \vec{p}} + \frac{\partial \hat{f}}{\partial \vec{p}} \cdot (\frac{e\vec{E}}{2\hbar} +$$
$$iBG^r(p,x,t) - B\hat{f} + \frac{J(p)}{4}\frac{\partial \vec{M}(x)}{\partial \vec{x}} \cdot \vec{\sigma} - iA\frac{\partial G^r}{\partial x} -$$
$$B\frac{\partial^2 \hat{f}}{\partial x \partial p}) - \frac{J(p)}{2i\hbar}\vec{M}(x) \cdot \vec{\sigma}\hat{f} + \frac{J(p)}{2i\hbar}\hat{f}\vec{M}(x) \cdot \vec{\sigma} + iAG^r\hat{f} -$$
$$iA\hat{f}G^r \quad = \quad -B\hat{f}\frac{\partial \hat{f}'}{\partial \vec{p}} + B\frac{\partial \hat{f}}{\partial \vec{p}}\hat{f}' + i\frac{\partial(n_i T_{pp})}{\partial x}\frac{\partial \hat{f}'}{\partial p} +$$
$$i\frac{\partial(n_i T_{pp})}{\partial x}\frac{\partial \hat{f}}{\partial p} - \frac{\partial(n_i T_{pp})}{\partial x}\frac{\partial G^r}{\partial p} + B\frac{\partial^2 \hat{f}}{\partial x \partial p}\frac{\partial \hat{f}'}{\partial x} - iA\frac{\partial \hat{f}}{\partial x}\frac{\partial G^r}{\partial x} +$$
$$iB\frac{\partial G^r}{\partial p}\frac{\partial^2 \hat{f}}{\partial x \partial p} - iB\frac{\partial^2 \hat{f}}{\partial^2 p}\frac{\partial G^r}{\partial x} - iB\frac{\partial G^r}{\partial x}\frac{\partial^2 \hat{f}}{\partial^2 p} + B\frac{\partial \hat{f}}{\partial x}\frac{\partial^2 \hat{f}}{\partial^2 p}$$
$$-A\frac{\partial \hat{f}}{\partial x}\frac{\partial \hat{f}'}{\partial p} + iB\frac{\partial^2 \hat{f}'}{\partial x \partial p} + iB\frac{\partial^2 \widehat{f'}}{\partial x \partial p} - \frac{\partial(n_i T_{pp})}{\partial p}\frac{\partial G^r}{\partial x} -$$
$$iB\frac{\partial^2 \hat{f}}{\partial x \partial p}\frac{\partial G^r}{\partial x}$$

## Appendix B